\documentclass[manuscript]{aastex}

\usepackage{amsbsy,amssymb,amsmath,bm}
\usepackage{graphicx,subfigure}
\usepackage{natbib}

\begin{document}

\title{The combined effect of precession and convection on the dynamo action}
\author{Xing Wei$^{1,2}$}
\affil{$^1$Institute of Natural Sciences and Department of Physics and Astronomy, Shanghai Jiao Tong University, Shanghai 200240, China \\ $^2$Princeton University Observatory, Princeton, NJ 08544, USA}
\email{xing.wei@sjtu.edu.cn, xingwei@astro.princeton.edu}

\begin{abstract}
To understand the generation of the Earth's and planetary magnetic fields, we investigate numerically the combined effect of precession and convection on the dynamo action in a spherical shell. The convection alone, the precession alone and the combined effect of convection and precession are studied at the low Ekman number at which the precessing flow is already unstable. The key result is that although the precession or convection alone is not strong to support the dynamo action the combined effect of precession and convection can support the dynamo action because of the resonance of precessional and convective instabilities. This result may interpret why the geodynamo maintains for such a long history compared to the Martian dynamo.

\end{abstract}

\maketitle

\section{Introduction}

Magnetic fields of astronomical bodies are generated by the dynamo action, namely the motion of electrically conducting fluid shears and twists magnetic field lines to create new field lines to offset magnetic diffusion. For the dynamo in the Earth's core, i.e. geodynamo, the thermal and compositional convection is believed to be the major power, in which the differential rotation and the helical motion combine to induce the dynamo action. The convection dynamo has been extensively studied since 1970's, e.g. \citet{busse}, \citet{hollerbach_review}, \citet{zhang_schubert}, \citet{glatzmaier_roberts}, \citet{jones} etc. On the other hand, \citet{bullard} discussed the possibility of precession driven geodynamo and \citet{malkus} pointed out that the flow instabilities driven by precession can power the geodynamo. The most recent work shows that the convection in the Earth's core may be not sufficiently strong for heat transfer as anticipated in the earlier studies \citep{olson}, which implies that the Earth's precession might be the major power for the geodynamo. Moreover, the magnetic records show that the geomagnetic dipole reversals are statistically correlated to the Earth's orbital eccentricity \citep{yamazaki}, which implies that the Earth's precession also plays an important role in the dipole reversals. These geophysical applications motivate the study of precession dynamo.

The precessing flow in the spheroidal geometry was studied by Poincar\'e \citeyearpar{poincare} for invisid fluid and by Busse \citeyearpar{busse-precession} for viscous fluid. Recently, the study of precessing flow attracts many attentions, e.g. the asymptotic study by \citet{zhang_chan_liao}, the numerical study by \citet{cebron,hollerbach_noir}, the experimental study by \citet{noir,lavorel_lebars,goto,lin}, etc. However, not many studies have been carried out for the precession dynamo because of the complex flow structure, e.g. the inertial waves spawn from the critical latitude, the thin internal shear layers, the triad resonance of instabilities, etc. \citep{kerswell,tilgner_review}. 

\citet{tilgner} carried out the first numerical calculation about precession dynamo in the spherical geometry with the spectral method. It was found that both the laminar precessing flow at high Ekman number and the unstable precessing flow at low Ekman number can induce dynamo. In the former the dynamo is powered by the poloidal flow arising from the Ekman layer, and in the latter by the instabilities of precessing flow. Later \citet{wu_roberts} carried out the finite difference calculation in the spheroidal geometry, and \citet{hansen} carried out the finite volume calculation in ellipsoidal geometry.

Then a question arises: what will be the combined effect of precession and convection on dynamo action? \citet{wei_tilgner} carried out the numerical calculation about the hydrodynamic interaction of precession and convection in a spherical shell. It was found that {\bf the two driving mechanisms for dynamo can destabilize each other}, namely the mutual interaction leads to a more unstable flow because of the resonance of the two instabilities, see the details in \citet{wei_tilgner}. Usually the flow instabilities favour the dynamo action, thus it seems plausible that the combined effect of precession and convection may facilitate the onset of dynamo and lead to the more efficient dynamo action.

In this paper, we extend the numerical calculations of the precession dynamo in \citet{tilgner} and of the precession-convection flow in \citet{wei_tilgner} to the precession-convection dynamo. We use the same numerical setup and code as in these two previous papers, i.e. the same linear stratification profile and the same precession angle $60^\circ$. In section 2 the mathematical equations are formulated and the numerical method is introduced. In section 3 the numerical results are shown and discussed. In section 4 this work is summarized, the possible applications to the geomagnetism, Martian magnetic field and the magnetic field in small bodies are briefly discussed, and the further study is pointed out.

\section{Equations}

The numerical setup is identical to that in \citet{tilgner} and \citet{wei_tilgner}. Suppose that we have a conducting fluid in a spherical shell with the aspect ratio $r_i/r_o$. The spherical shell spins at the rate $\Omega_s$ about its symmetric axis (the $z$ axis) and precesses at the rate $\Omega_p$ about an inclined axis with the angle $\beta$ to the $z$ axis. In the frame attached to the boundary, the unit vector of precession axis is expressed in the Cartesian coordinate system ($x,y,z$) as
\begin{equation}
\hat{\bm\Omega}_p=\sin\beta\cos t\hat{\bm x}-\sin\beta\sin t\hat{\bm y}+\cos\beta\hat{\bm z},
\end{equation}
where hat denotes unit vector. In the meanwhile, we impose a background temperature $T_b$ and assume it a linear profile
\begin{equation}
T_b=\frac{T_o-T_i}{d}(r-r_o)+T_o,
\end{equation}
where $T_o$ and $T_i$ are the temperature respectively at $r_o$ and $r_i$ and $d$ the thickness of the spherical shell. The temperature gradient $(T_o-T_i)/d$ is negative (i.e. unstable stratification) for convection. This linear temperature profile is maintained by a heat source inversely proportional to radius. In addition to the linear profile we can assume other profiles maintained by different heat sources. We choose the linear profile because it is simple in terms of numerics, namely in the physical space the grids near the boundaries are not required to be dense such that the Chebyshev collocation points we use in the radial direction are sufficiently dense to resolve the linear profile.

We make use of the Boussinesq approximation that the density variation is considered only in the buoyancy force and proportional to temperature deviation $\Theta=T-T_b$. Then the dimensionless Navier-Stokes equation in the frame attached to the boundary reads
\begin{equation}\label{ns}
\frac{\partial\bm u}{\partial t}+\bm u\cdot\bm\nabla\bm u=-\bm\nabla\Phi+Ek\nabla^2\bm u+2\bm u\times(\hat{\bm z}+Po\hat{\bm\Omega}_p)+Po(\hat{\bm z}\times\hat{\bm\Omega}_p)\times\bm r+\widetilde{Ra}\Theta\bm r+(\bm\nabla\times\bm B)\times\bm B,
\end{equation}
where all the curl-free terms are absorbed into the total potential $\Phi$. On the right-hand-side of \eqref{ns}, the second term is viscous force, the third is the Coriolis force due to global rotation, the fourth is the Poincar\'e force due to precession and it drives the precessing flow, the fifth is the buoyancy force due to stratification, and the last is the Lorentz force due to magnetic field. The dimensionless temperature deviation equation reads
\begin{equation}\label{temperature}
\frac{\partial\Theta}{\partial t}+\bm u\cdot\bm\nabla\Theta=\frac{Ek}{Pr}\nabla^2\Theta+u_r,
\end{equation}
On the right-hand-side of \eqref{temperature}, the inhomogeneous term $u_r$ derived from the advection term $\bm u\cdot\bm\nabla T_b$ causes the temperature deviation. The dimensionless magnetic induction equation reads
\begin{equation}\label{induction}
\frac{\partial\bm B}{\partial t}=\bm\nabla\times(\bm u\times\bm B)+\frac{Ek}{Pm}\nabla^2\bm B.
\end{equation}

In the above dimensionless equations (\ref{ns}), (\ref{temperature}) and (\ref{induction}), the normalization is as follows. Length is normalized with the shell thickness $d$, time with the inverse of rotation rate $\Omega_s^{-1}$, velocity with $\Omega_sd$, temperature deviation with $T_i-T_o$, and magnetic field with $\sqrt{\rho\mu}\Omega_sd$ (where $\rho$ is the fluid density and $\mu$ the magnetic permeability). There are five dimensionless parameters governing the system. The Ekman number $Ek=\nu/(\Omega_sd^2)$ measures the ratio of viscous time scale to spin time scale, the Poincar\'e number $Po=\Omega_p/\Omega_s$ measures the ratio of precession rate to spin rate, the rotational Rayleigh number $\widetilde{Ra}=\alpha g_o(T_i-T_o)/(\Omega_s^2r_o)$ (where $\alpha$ is the thermal expansion coefficient and $g_o$ the gravitational acceleration at $r_o$) measures the square of ratio of buoyancy frequency to spin rate, the Prandtl number $Pr=\nu/\kappa$ measures the ratio of viscosity to thermal diffusivity and the magnetic Prandtl $Pm=\nu/\eta$ measures the ratio of viscosity to magnetic diffusivity. It should be noted that what we use to measure the strength of convection is the rotational Rayleigh number but not the conventional Rayleigh number $Ra=\alpha g_o(T_i-T_o)r_o^3/\nu\kappa$. They are related through
\begin{equation}\label{ra}
Ra=\frac{\alpha g_o(T_i-T_o)d^3}{\nu\kappa}=\frac{\alpha g_o(T_i-T_o)}{\Omega_s^2r_o}\,\frac{\Omega_s^2d^4}{\nu^2}\,\frac{\nu}{\kappa}\,\frac{r_o}{d}=\widetilde{Ra}\,Ek^{-2}\,Pr\,\frac{r_o}{d}.
\end{equation}
The aspect ratio is given to be $0.1$ to minimize the effect of inner core and the precession angle $\beta$ to be $60^\circ$ as in \citet{tilgner} and \citet{wei_tilgner} such that precession has a noticeable effect. 

The velocity boundary condition is no-slip $\bm u=\bm 0$ at the outer boundary (precession in spherical geometry couples the fluid motion and the boundary motion through viscosity and therefore the no-slip outer boundary condition is necessary to drive the precessing flow) and stress-free at the inner boundary to approximate a full sphere. The boundary condition for temperature deviation is homogeneous $\Theta=0$. The magnetic boundary condition is insulating, namely magnetic field at the boundaries matches a potential field for the exterior regions of $r>r_o$ and $r<r_i$. The initial values of flow, temperature deviation and magnetic field are given to be small values.

The equations are numerically solved in spherical coordinate system ($r,\theta,\phi$) with a pseudo-spectral code \citep{tilgner_numerics} which was used in \citet{tilgner} and \citet{wei_tilgner}. The toroidal-poloidal decomposition is used to take into account the solenoidal property of fluid velocity and magnetic field. All the functions are expanded with the spherical harmonics on the spherical surface and with the Chebyshev polynomials in the radial direction. The semi-implicit scheme is employed for time stepping, using an Adams-Bashforth scheme for the nonlinear terms and a Crank-Nicolson scheme for the diffusive terms. Resolution as high as $128^3$ is used and the resolutions are checked as in \citet{tilgner} and \citet{wei_tilgner}. To identify a successful dynamo, we integrate the MHD equations until the magnetic energy grows to a noticeable value and maintains for a long period without the tendency to decay.

\section{Results}

As discussed in \citet{tilgner}, both the laminar precessing flow at high Ekman number and the unstable precessing flow at low Ekman number can induce the dynamo action. In the Earth's core, the Ekman number is very low, of the order of $10^{-15}$, and the precessing flow at such a low Ekman number is unstable. Therefore, we study the Ekman number $Ek=3\times10^{-4}$ at which the precessing flow is already unstable \citep{tilgner,wei_tilgner}. One may argue that this Ekman number is not low enough, e.g. $10^{-6}$ used in some large-scale simulations for convection dynamo. It should be noted that the purpose of our study is to embark the investigation of the combined effect of precession and convection on dynamo but neither to push the parameter towards the real Earth nor to scan the parameter space for a systematic study. It is clear that at a lower Ekman number the precessing flow has more complex structure and exhibits higher azimuthal wavenumbers, which favours the dynamo action. At $Ek=3\times10^{-4}$ the precessing flow is already unstable and the important physical ingredient that {\bf an unstable precessing flow favours the dynamo action} is already involved. So we do not attempt to push the Ekman number to smaller values due to the limitation of our computational facility.

We calculate step-by-step the precession dynamo, the convection dynamo, and the precession-convection dynamo. The Prandtl number $Pr$ is fixed to be $1$. It is known that a higher magnetic Prandtl number $Pm$ facilitates the onset of dynamo, see Figure 1 in \citet{christensen} or Figure 8 in \citet{jones}, and it is fixed to be $2$ which is above the critical $Pm$. We then vary $Po$ and $\widetilde{Ra}$ to search dynamo. Because the Earth's precession is retrograde, $Po$ is given to be negative.

\begin{figure}
\centering
\subfigure[]{\includegraphics[scale=0.3]{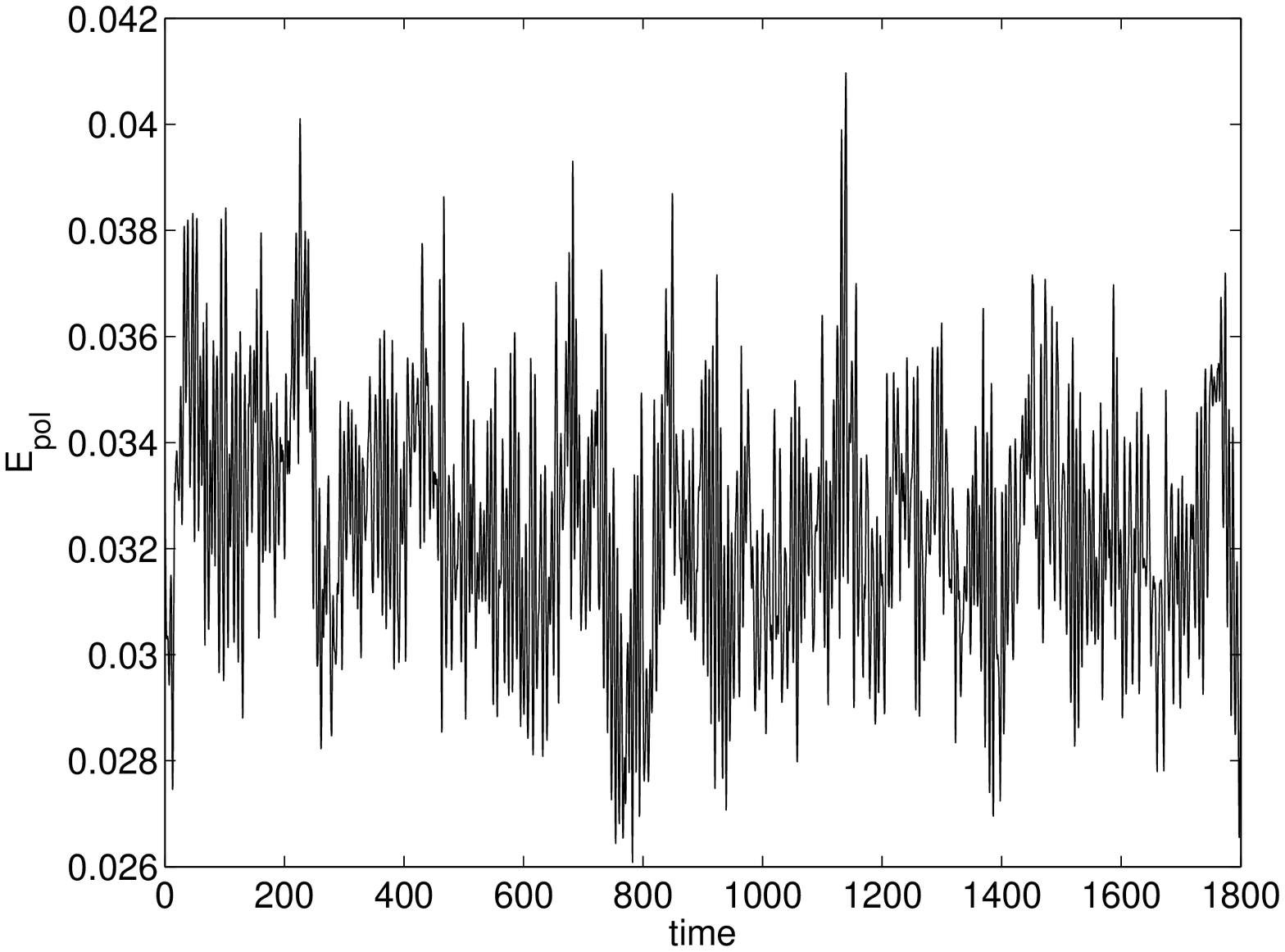}\label{fig1a}}
\subfigure[]{\includegraphics[scale=0.3]{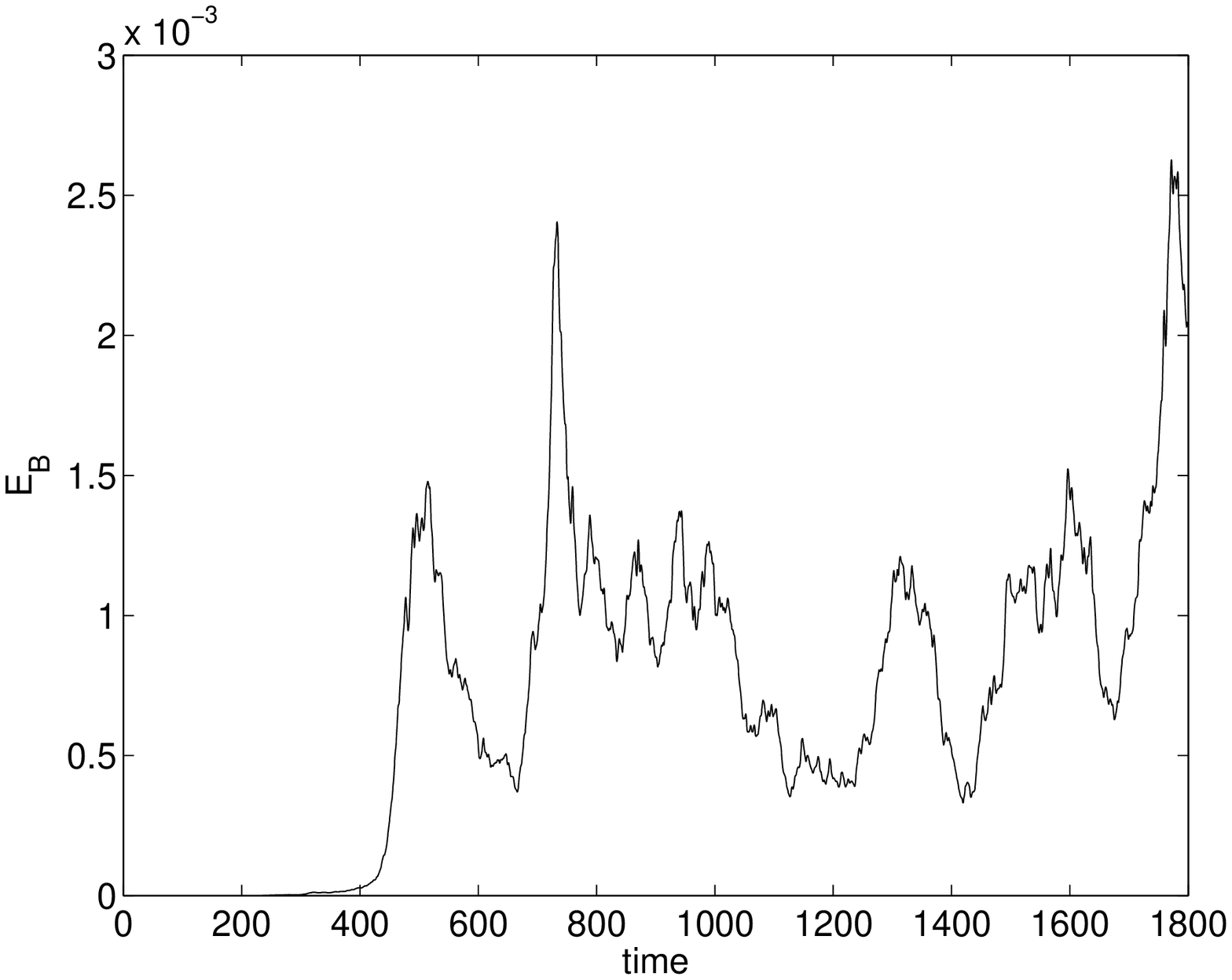}\label{fig1b}}
\caption{The precession dynamo at $Po=-0.3$. The time evolution of poloidal kinetic energy (a) and magnetic energy (b).}\label{fig1}
\end{figure}

Firstly we study the precssion dynamo. Before calculating the precession dynamo, we calculate the hydrodynamic precessing flow. These numerical calculations give the fluid rotation vector to be consistent with Busse's solution derived from the Ekman-layer asymptotic calculation \citep{busse}. We do not repeat to show the hydrodynamic results in this paper, which have been already discussed in detail in \citet{tilgner} and \citet{wei_tilgner} using the same numerical setup and code. Now we vary $Po$ to calculate the nonlinear dynamo equations at $Ek=3\times10^{-4}$, $Pr=1$ and $Pm=2$ to search the critical $|Po|$ for the onset of dynamo. We increase $|Po|$ by a step of $0.1$. It is found that magnetic energy eventually decays at $Po=-0.2$, but grows and eventually saturates at $Po=-0.3$. Therefore, the critical $|Po|$ for the precession dynamo is between $0.2$ and $0.3$. At such the low Ekman number, the precssional instabilities develop such that the symmetry of laminar precessing flow about the centre ($r=0$) breaks, and so the instabilities can be measured by the kinetic energy of anti-symmetric component of flow $\bm u_a=[\bm u(\bm r)+\bm u(-\bm r)]/2$ \citep{tilgner,wei_tilgner}. It should be noted that the precessional instabilities contain both anti-symmetric and symmetric components, but the laminar precessing flow contains only the symmetric component, and so the non-zero anti-symmetric component indicates the precessional instabilities and the energy of the anti-symmetric component measures the strength of the precessional instabilites. Figure \ref{fig1} shows the time evolution of the precession dynamo at $Po=-0.3$, which is consistent with the result in \citet{tilgner}. The ratio of anti-symmetric kinetic energy to total kinetic energy $E_a/E_{kin}$ is not negligible and its time average is $5.34\times10^{-3}$, which indicates that the precessing flow is unstable. The poloidal flow is important for the $\alpha$ effect in dynamo action, i.e. twisting field lines. Figure \ref{fig1a} shows the time evolution of poloidal kinetic energy. Its time-average is listed in Table \ref{table}. As in \citet{tilgner} we define the magnetic Reynolds number $Rm$ with the dimensionless mean poloidal flow $u_{\rm pol}=\sqrt{2E_{\rm pol}/V}$ (where $E_{\rm pol}$ is the poloidal energy of flow and $V$ is the fluid volume) to be $Rm=u_{\rm pol}Pm/Ek$. $Rm$ is 707 at $|Po|=0.3$ (Table \ref{table}) for a successful dynamo, but 694 at $|Po|=0.2$ for a failed dynamo. The dominant azimuthal mode of precessing flow is $m=1$, i.e. the spin-over mode \citep{greenspan,tilgner_review}. Figure \ref{fig1b} shows the time evolution of magnetic energy. Comparison between Figures \ref{fig1a} and \ref{fig1b} indicates that flow fluctuates on a small scale whereas magnetic field varies on a large time scale. The time-average of magnetic energy $E_{\rm B}$ is also listed in Table \ref{table}.

\begin{figure}
\centering
\subfigure[]{\includegraphics[scale=0.3]{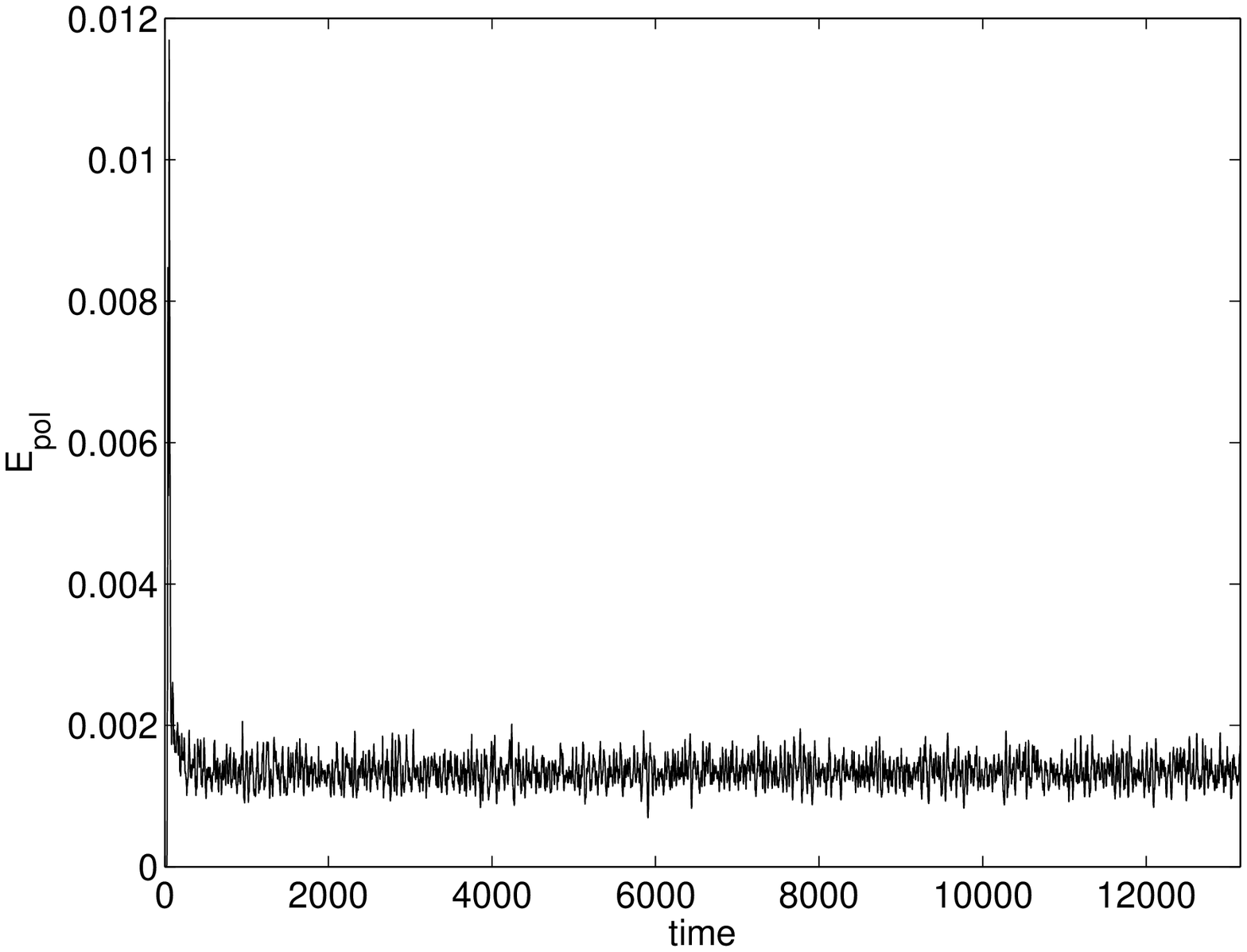}\label{fig2a}}
\subfigure[]{\includegraphics[scale=0.3]{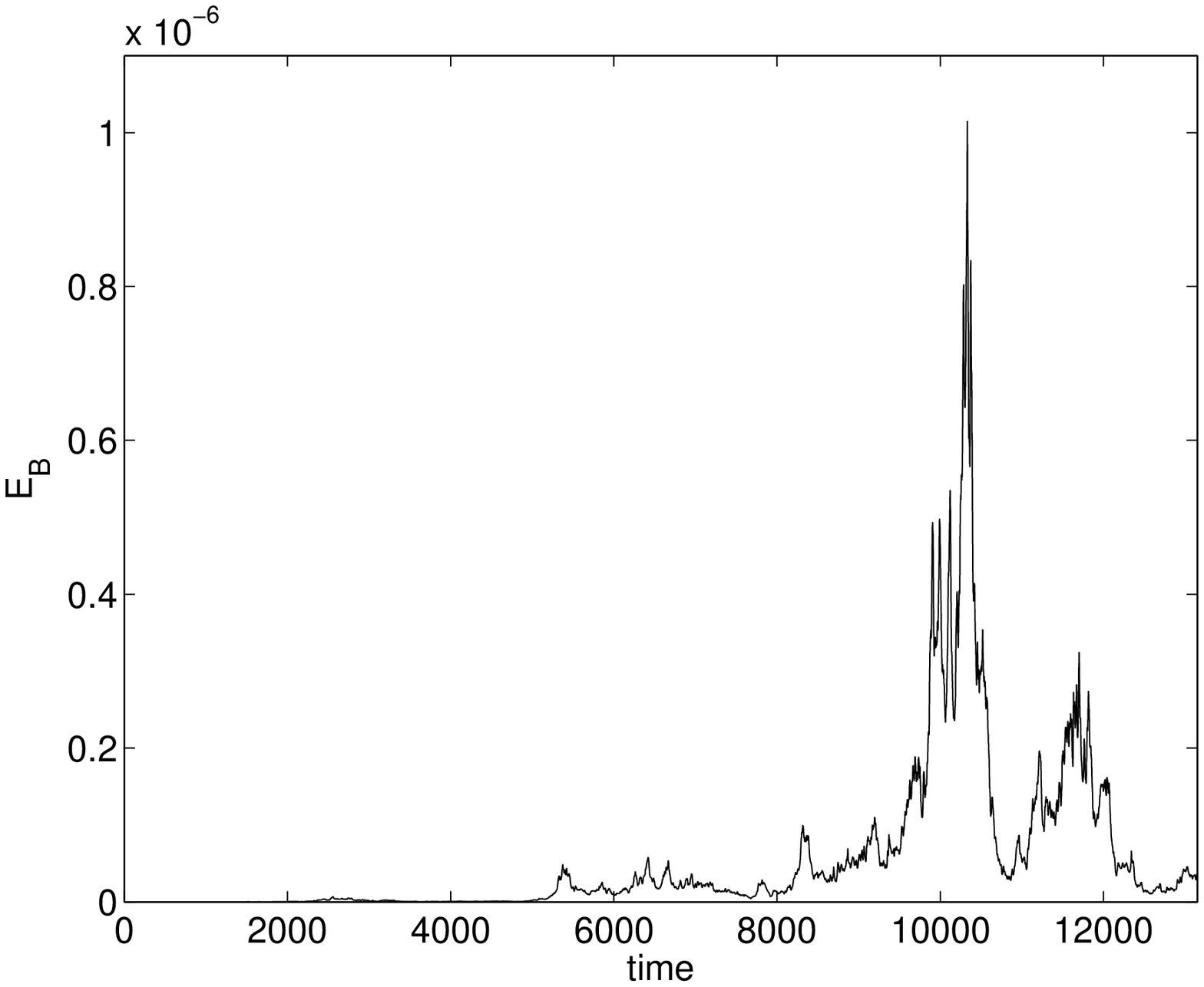}\label{fig2b}}
\caption{The convection dynamo at $\widetilde{Ra}=0.6$. The time evolution of poloidal kinetic energy (a) and magnetic energy (b).}\label{fig2}
\end{figure}

Next we study the convection dynamo. Similar to the precession dynamo, we vary $\widetilde{Ra}$ to search the onset of the convection dynamo. It is found that the critical $\widetilde{Ra}$ is between $0.5$ and $0.6$, i.e. magnetic energy decays at $\widetilde{Ra}=0.5$ but grows and saturates at $\widetilde{Ra}=0.6$. Figure \ref{fig2} shows the time evolution of the convection dynamo at $\widetilde{Ra}=0.6$. We need to emphasize that although the difference of $\widetilde{Ra}$ is only 0.1 the conventional $Ra$ translated through equation \eqref{ra} is more than one million! (see Table 1). In a convective flow, Nusselt number $Nu$ is used to measure the ratio of the total heat flux to the thermal conduction. The time average of $Nu$ at the outer boundary is 1.83 (Table \ref{table}) which indicates a strong convective motion ($Nu$ at the inner boundary can be deduced from its value at the outer boundary through equation (4.2) in \citet{wei_tilgner}). Figure \ref{fig2a} shows the time evolution of poloidal kinetic energy. The poloidal kinetic energy of the convection dynamo is much lower than that of the precession dynamo and thus $Rm=144$ for the convection dynamo is lower than $Rm=707$ for the precession dynamo (Table \ref{table}). As we have discussed in the last paragraph, at $Rm=694$ the precessing flow cannot maintain a dynamo, but at $Rm=144$ the convective flow can. In this sense, convection is more efficient for the onset of dynamo than precession. We need to point out that this conclusion is valid only at this Ekman number. At a smaller Ekman number, the precessing flow is more complex which favours the onset of dynamo, and this conclusion may not hold any longer (we leave the large-scale simulations at smaller Ekman numbers for the other researchers who will be interested in the result of this work). The comparison between Figures \ref{fig1a} and \ref{fig2a} suggests that not only the mean poloidal kinetic energy but also the fluctuation amplitude of precessing flow are much higher than those of convective flow. This implies that the precessional instabilities are more vigorous than the convective instabilities in the two successful dynamos. The dominant azimuthal mode in the convective flow is $m=3$ (Table \ref{table}), indicating a shorter length scale than the dominant spin-over mode $m=1$ in the precessing flow. Figure \ref{fig2b} shows the time evolution of magnetic energy. Compared to Figure \ref{fig1b}, the magnetic energy of the convection dynamo becomes noticeable at time $\approx5000$, which is much later than time $\approx400$ of the precession dynamo. It is not surprising that the magnetic energy of the convection dynamo is also much lower than of the precession dynamo (Table \ref{table}) because of the lower $Rm$ of the former.

\begin{figure}
\centering
\subfigure[]{\includegraphics[scale=0.3]{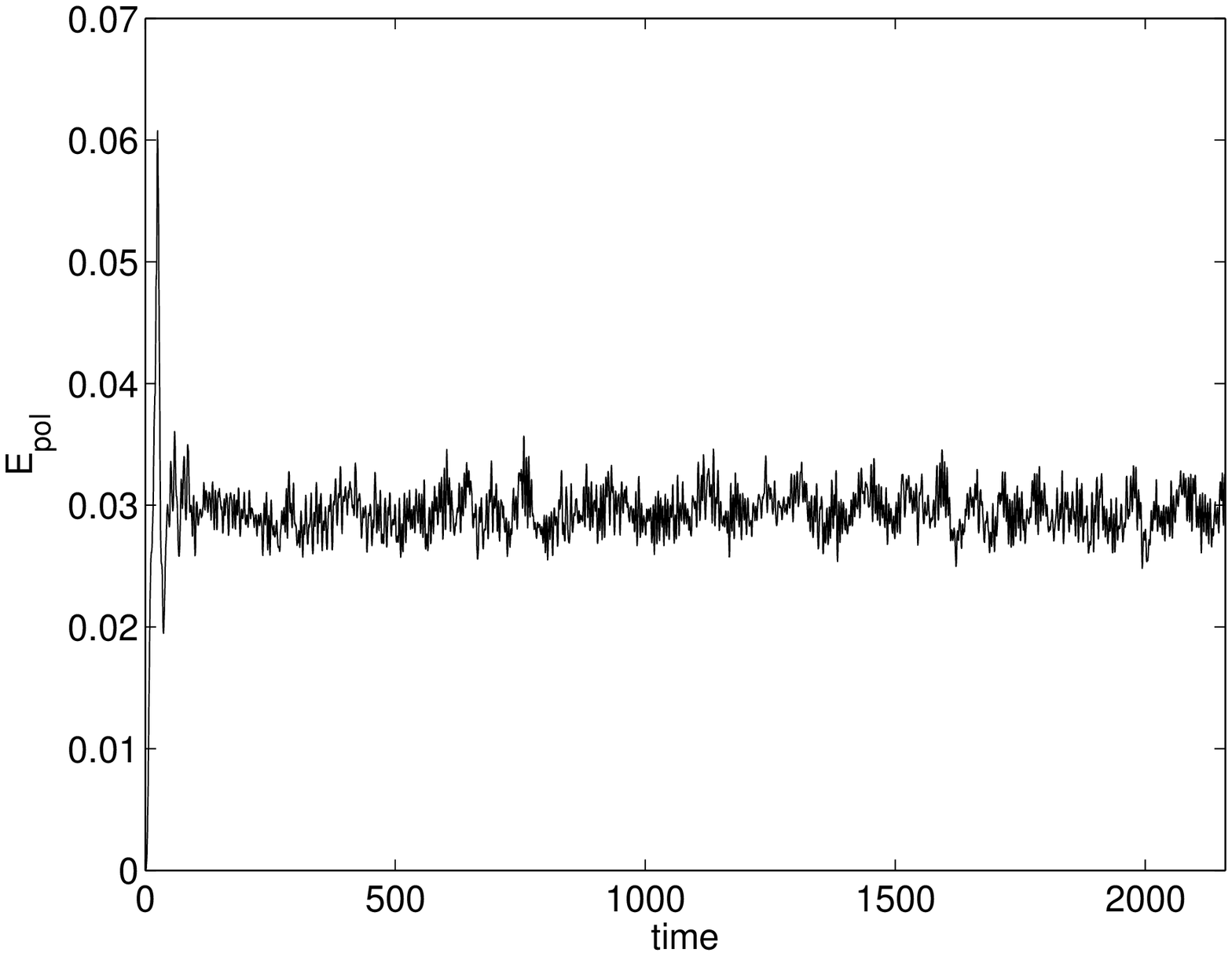}\label{fig3a}}
\subfigure[]{\includegraphics[scale=0.3]{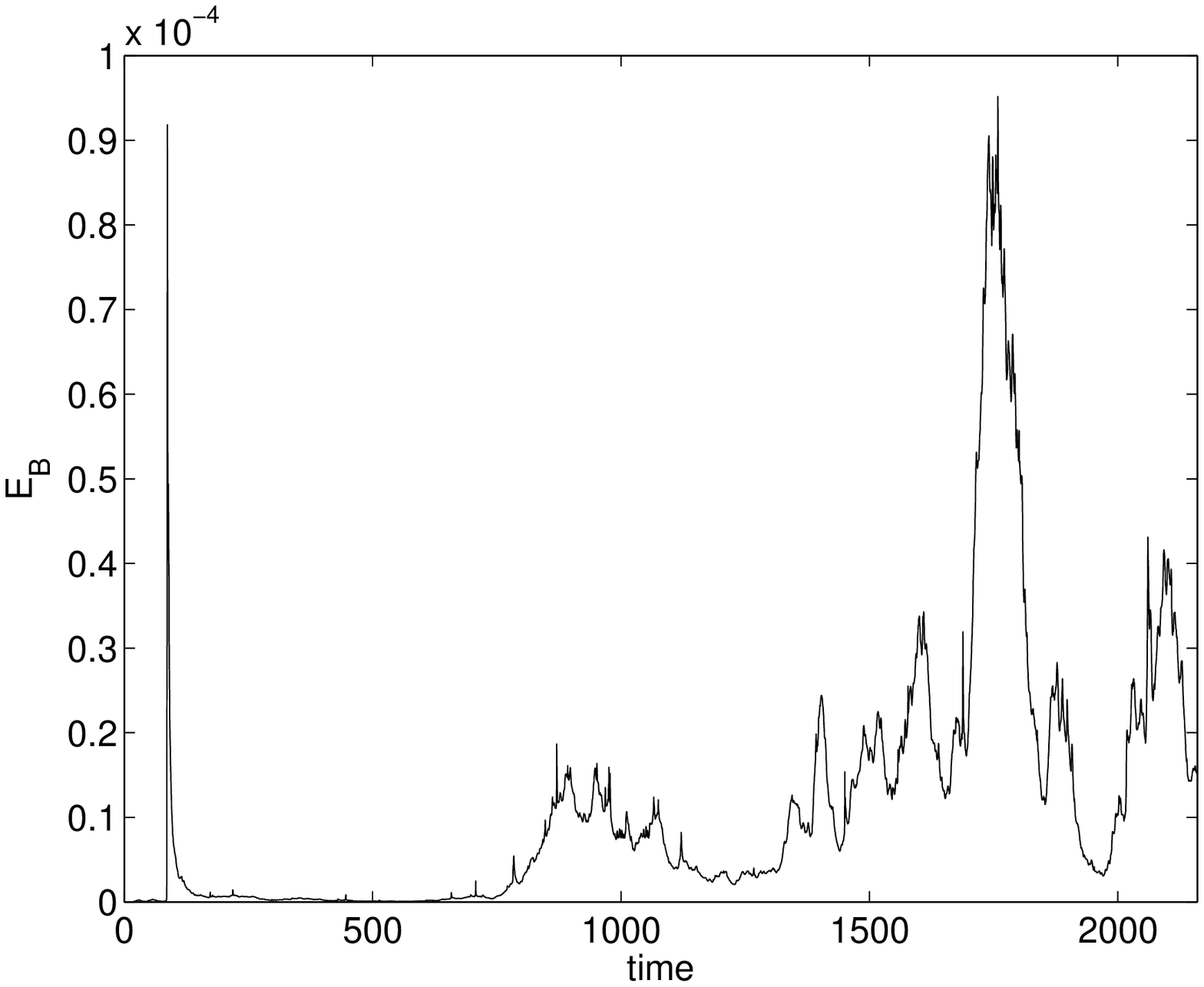}\label{fig3b}}
\caption{The precession-convection dynamo at $Po=-0.2$ and $\widetilde{Ra}=0.5$. The time evolution of poloidal kinetic energy (a) and magnetic energy (b).}\label{fig3}
\end{figure}

After studying the dynamos driven by precession alone and by convection alone, we study the combined effect of precession and convection. In the above two dynamos, the dynamo driven by precession alone cannot be maintained at $Po=-0.2$ and the dynamo driven by convection alone cannot be maintained at $\widetilde{Ra}=0.5$. We test whether the preceission-convection dynamo can be maintained at $Po=-0.2$ and $\widetilde{Ra}=0.5$. {\bf This dynamo works!} It indicates that the combined effect of precession and convection can indeed facilitate the onset of dynamo. As we discussed, the conventional $Ra$ differs by more than one million, which indicates that precession greatly helps the onset of convection dynamo. Moreover, it is interesting that the $Rm=677$ of precession-convection dynamo is lower than the $Rm=694$ of the failed precession dynamo, as shown in Table 1. This suggests that the combination of precession and convection has some non-trivial effect and it triggers the dynamo action at a lower $Rm$ of pure precession dynamo. This non-trivial effect probably arises from {\bf the resonance of precessional instability and convective instability}. Figure \ref{fig3} shows the time evolution of this precession-convection dynamo and its time-averaged values are listed in Table \ref{table}. Although $\widetilde{Ra}=0.5$ of the precession-convection dynamo is lower than $\widetilde{Ra}=0.6$ of the convection dynamo, $Nu$ of the former is higher (Table \ref{table}). This is because the poloidal flow driven by precession contributes more to heat transfer than convection, i.e. $E_{\rm pol}=3.11\times10^{-2}$ of the precessing flow at $Po=-0.2$ is already much higher than $E_{\rm pol}=1.33\times10^{-3}$ of the convective flow at $\widetilde{Ra}=0.6$. The dominant azimuthal mode is $m=1$ (Table \ref{table}), which indicates that the precession-convection flow is more precessing than convective. $E_{\rm pol}$ and $Rm$ of the precession-convection dynamo are a little lower than those of precession dynamo but much higher than those of convection dynamo (Table \ref{table}), which again suggests that the precession-convection flow is more precessing. The magnetic energy becomes noticeable at time $\approx700$ (Figure \ref{fig3b}, which is a little later than time $\approx400$ of the precession dynamo but much earlier than time $\approx5000$ of the convection dynamo.

\begin{table}
\centering
\begin{tabular}{c|ccccccc}
($Po$, $\widetilde{Ra}$) & $Ra$ & $Nu$ & $E_{\rm pol}$ & $Rm$ & $m$ & $E_{\rm B}$ \\
\hline
(-0.2, 0) & 0 & n/a & $3.11\times10^{-2}$ & 694 & 1 & {failed dynamo} \\
(-0.3, 0) & 0 & n/a & $3.23\times10^{-2}$ & 707 & 1 & $9.44\times10^{-4}$ \\
(0, 0.5) & $6.17\times10^6$ & 1.78 & $1.15\times10^{-3}$ & 133 & 3 & {failed dynamo} \\
(0, 0.6) & $7.41\times10^6$ & 1.83 & $1.33\times10^{-3}$ & 144 & 3 & $2.11\times10^{-7}$  \\
(-0.2, 0.5) & $6.17\times10^6$ & 2.20 & $2.96\times10^{-2}$ & 677 & 1 & $1.78\times10^{-5}$ 
\end{tabular}
\caption{The successful and failed dynamos at different $Po$ and $\widetilde{Ra}$. For the successful dynamos, the conventional Rayleigh number, the Nusselt number, poloidal kinetic energy, dominant azimuthal mode of flow and magnetic energy are shown. The values are taken for time-average in the statistically steady stage.}\label{table}
\end{table}

\begin{figure}
\centering
\subfigure[]{\includegraphics[scale=0.4]{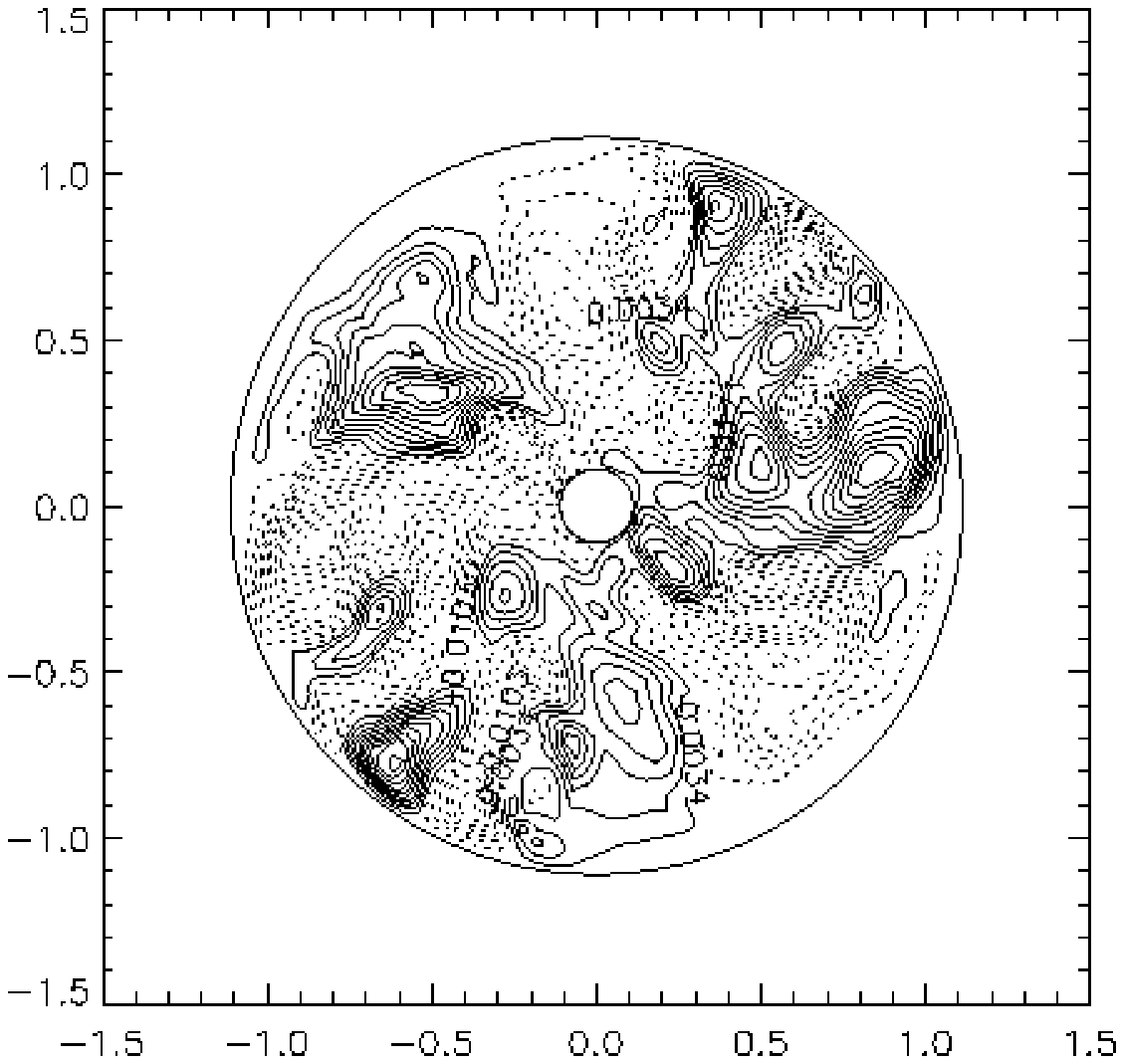}\label{fig4a}}
\subfigure[]{\includegraphics[scale=0.4]{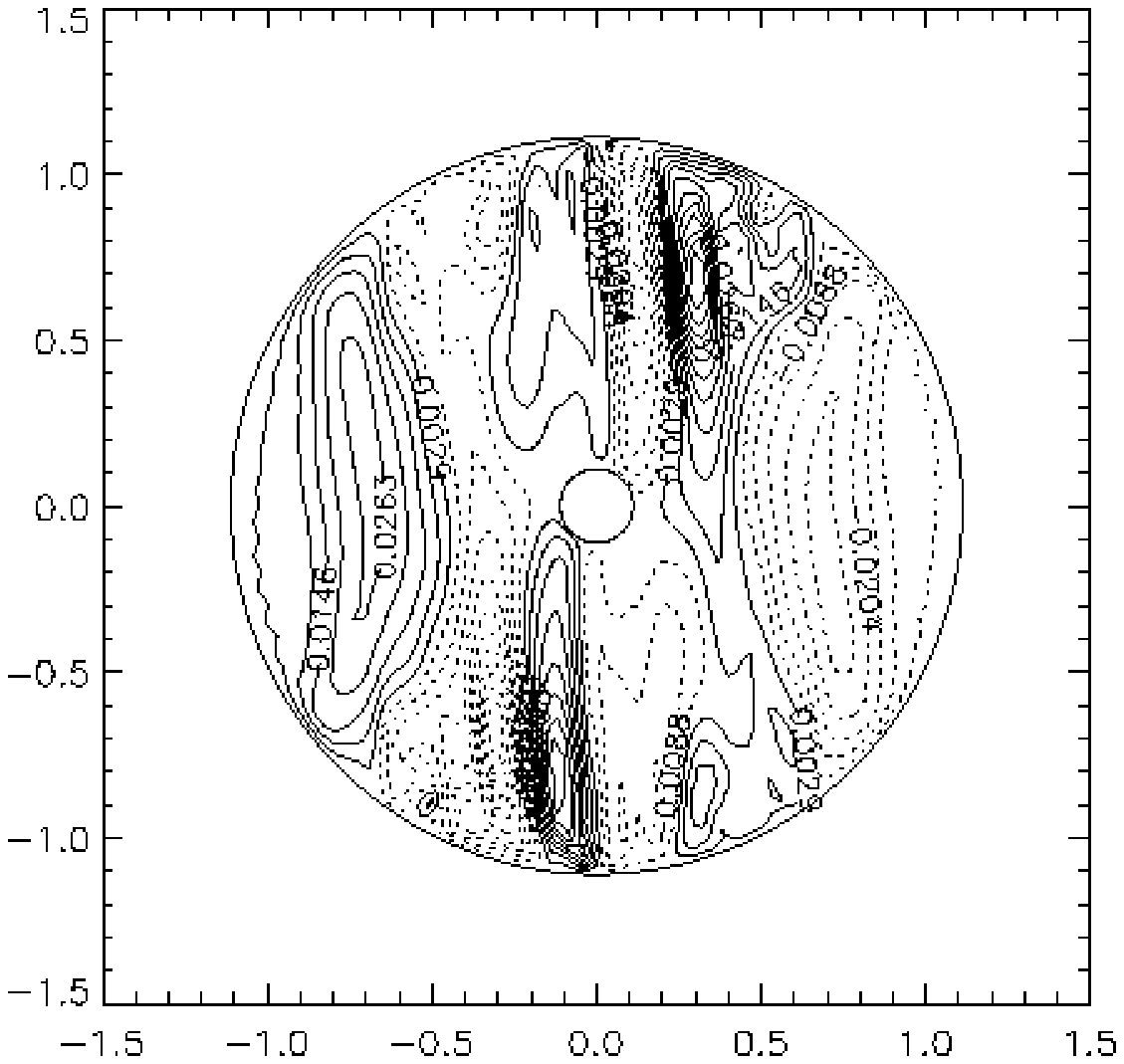}\label{fig4b}}
\subfigure[]{\includegraphics[scale=0.4]{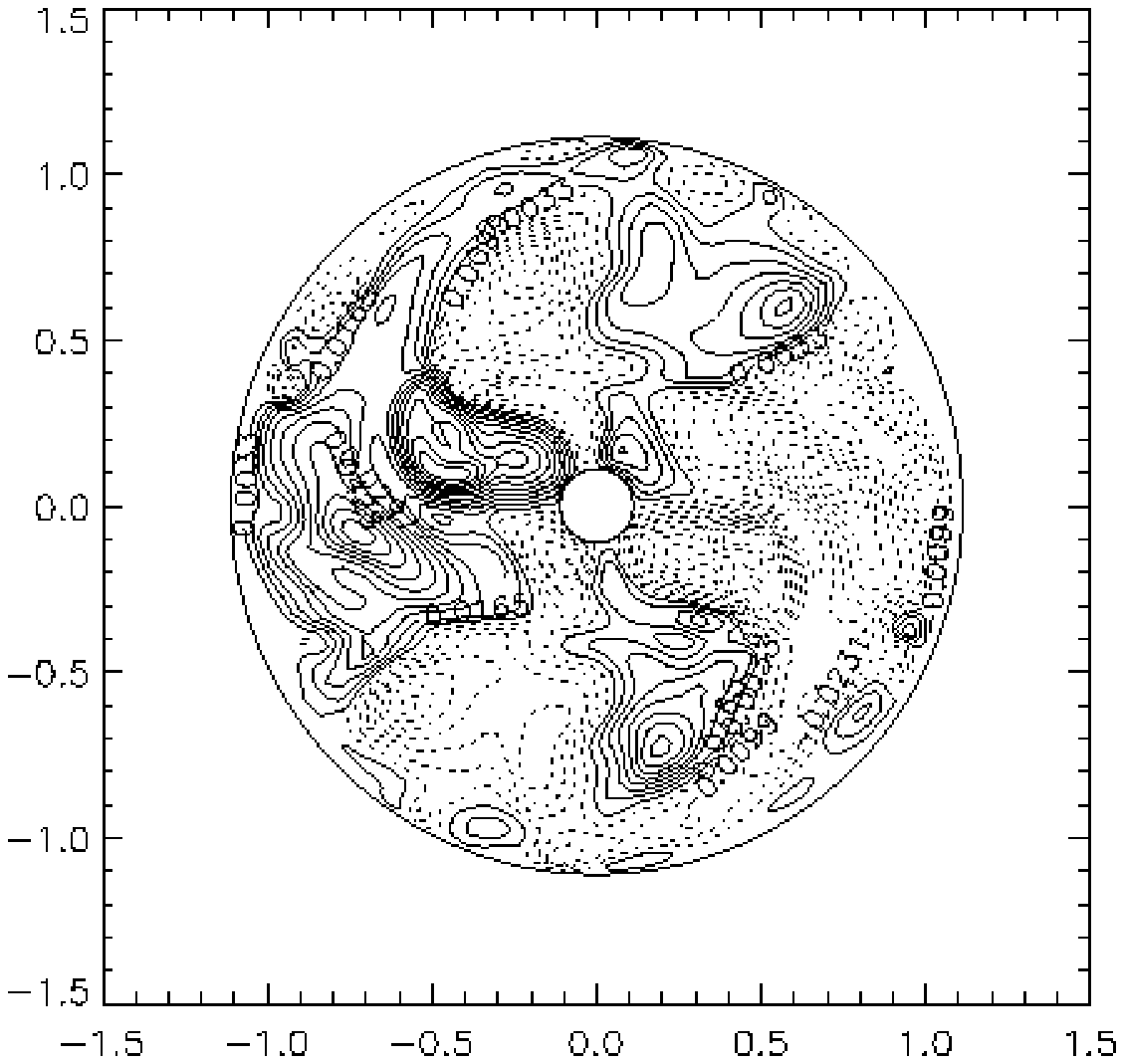}\label{fig4c}}
\caption{The contours of the radial velocity in the meridional plane at $\phi=180^\circ$. (a) The precession dynamo at $Po=-0.3$. (b) The convection dynamo at $\widetilde{Ra}=0.6$. (c) The precession-convection dynamo at ($Po=-0.2$, $\widetilde{Ra}=0.5$). Solid lines denote positive and dotted lines negative. They are all snapshots when the dynamos saturate.}\label{fig4}
\end{figure}

To end this section, we discuss the flow patterns in the different dynamos. Figure \ref{fig4} shows the contours of the radial velocity in the meridional plane in the precession, convection and precession-convection dynamos. The flow of the convection dynamo exhibits the columnar structure at such the low $Ek$. But the flow of the precession dynamo seems chaotic because, as discussed, the precessional instabilities are more vigorous than the convective instabilities. The flow of the precession-convection dynamo is similar to that of the precession dynamo and has the more complex structure than the flow of the convection dynamo. This also explains why the combined effect favours for the dynamo action. It is because the precession-convection flow tends to be chaotic and have the complex structure which favours the dynamo action.

\section{Summary}
Through our numerical calculations we know that the combined effect of precession and convection favours the dynamo action. {\bf Although the precession alone or the convection alone is not strong to support the dynamo action, the combined precession-convection dynamo works}. The reason is that the combined effect tends to make the flow more unstable and the more complex flow structure emerges, which favours the dynamo action. Then we may have a tentative point. After a long history, the heat flux in the Earth's fluid core becomes weaker and weaker and at some time the convection is not powerful to support the geodynamo, e.g. \citep{olson}, but the geodynamo can still be maintained because the precession provides the energy. This could have already occurred in the Earth's early history when the fluid core was too small to support the geodynamo. This could be occurring in the Earth's core. This could be to occur in the future because the heat flux in the Earth's fluid core diminishes and will not support the geodynamo. As comparison, the Martian dynamo terminates because the precession of Mars is not as strong as that of the Earth. When the convection in the Martian fluid core stopped the Martian dynamo cannot be maintained by the weak Martian precession. It should be clarified that this is our tentative conjecture and needs more observational evidences to support or deny.

In addition to the Earth's magnetic field, the result of this work can be extended to the magnetic fields of small bodies. \citet{wei-arlt-tilgner} studied the dynamo action in small bodies driven by collisions. Precession can be considered as continuous collisions when the collision frequency is close to infinitesimal. In the presence of both collision and convection, it is plausible that the dynamo due to collisions tends to be driven more easily than collision or convection alone.

There is some further work that we leave for the researchers who will be interested. The Ekman number in this work is not very small, although sufficiently small for the onset of precessional instabilities. Our work simply initiates the study about the combined effect of precession and convection. As we know, the precessing flow structure at lower Ekman numbers will be complex and instabilities will prefer higher azimuthal modes. Therefore the low Ekman regime is necessary to investigate. Another further study is the geometry. In this work we study the spherical dynamo. The Earth's core is not spherical but spheroidal. The pressure torque in spheroidal geometry can enhance the coupling between fluid and boundary motions \citep{tilgner_review}, and moreover, the elliptical instability in spheroidal geometry, an instability of a two-dimensional flow with elliptical streamlines leading to a three-dimensional flow \citep{kerswell_elliptical,tilgner_review,zhang_chan_liao}, can occur. Both the pressure torque and the elliptical instability can facilitate the dynamo action. Therefore the spheroidal geometry is also necessary to investigate.

\section*{Acknowledgments}
This work was initiated in Princeton and completed in Shanghai. Prof. Andreas Tilgner provided me his code. Prof. Andreas Tilgner and Prof. Keke Zhang gave me valuable suggestions about this work. This work was supported by the National Science Foundation’s Center for Magnetic Self-Organization under grant PHY-0821899 and the startup grant WF220441903 of Shanghai Jiao Tong University.

\bibliographystyle{apj}
\bibliography{paper}

\label{lastpage}
\end{document}